\documentclass[letter,bibyear]{aa} % if the references are not structured
\usepackage{epsfig}
\usepackage{amsmath}
\usepackage{subfigure}
\usepackage{natbib}
\usepackage{multirow}
\usepackage{color}
\usepackage{longtable}
\usepackage{graphicx}
\usepackage{epstopdf}
\usepackage{booktabs}
\usepackage{txfonts}
\newcommand{\jmst}{J.~Mol.~Struct.}   % Journal of Molecular Structure

\begin{document}

\title{Discovery of HC$_3$O$^+$ in space: The chemistry of O-bearing species in TMC-1\thanks{Based on observations carried out
with the Yebes 40m telescope (projects 19A003 and
20A014) and the Institut de Radioastronomie Millim\'etrique (IRAM) 30m telescope. The 40m radio telescope at Yebes Observatory is
 operated by the Spanish Geographic Institute
  (IGN, Ministerio de Transportes, Movilidad y Agenda Urbana);
  IRAM is supported by INSU/CNRS (France), MPG (Germany), and IGN (Spain).}}

\author{
J.~Cernicharo\inst{1},
N.~Marcelino\inst{1},
M.~Ag\'undez\inst{1},
Y.~Endo\inst{2},
C.~Cabezas\inst{1},
C.~Berm\'udez\inst{1},
B.~Tercero\inst{3,4}, and
P.~de Vicente\inst{4}
}

\institute{Grupo de Astrof\'isica Molecular, Instituto de F\'isica Fundamental (IFF-CSIC), C/ Serrano 121, 28006 Madrid, Spain.
\email: jose.cernicharo@csic.es
\and Department of Applied Chemistry, Science Building II, National Chiao Tung University, 1001 Ta-Hsueh Rd., Hsinchu 30010, Taiwan
\and Observatorio Astron\'omico Nacional (IGN), C/ Alfonso XII, 3, 28014, Madrid, Spain.
\and Centro de Desarrollos Tecnol\'ogicos, Observatorio de Yebes (IGN), 19141 Yebes, Guadalajara, Spain.
}

\date{Received; accepted}

\abstract{
Using the Yebes 40m and IRAM 30m radio telescopes, we detected a series
of harmonically related lines with a rotational constant $B_0$=4460.590$\pm$0.001 MHz
and a distortion constant $D_0$=0.511$\pm$0.005 kHz towards the cold dense core TMC-1.
High-level-of-theory \emph{ab initio} calculations indicate that the best possible
candidate is protonated tricarbon monoxide, HC$_3$O$^+$. We have succeeded in producing
this species in the laboratory and observed its $J_u$-$J_l$ = 2-1 and 3-2 rotational transitions. Hence,
we report the discovery of HC$_3$O$^+$ in space based on our observations, theoretical calculations,
and laboratory experiments. We derive an abundance ratio
$N$(C$_3$O)/$N$(HC$_3$O$^+$)$\sim$7. The high abundance of the protonated form of C$_3$O is
due to the high proton affinity of the neutral species.
The chemistry of O-bearing species is modelled, and predictions are compared to the
derived abundances from our data for the most prominent O-bearing species in TMC-1.
}


\titlerunning{HC$_3$O$^+$ in TMC-1}
\authorrunning{Cernicharo et al.}

\maketitle

\section{Introduction}
The cold dark core TMC-1 presents an interesting carbon-rich chemistry that leads to
the formation of long neutral carbon-chain radicals and their anions
(see \citealt{Cernicharo2020a} and references therein). Cyanopolyynes, which are stable molecules,
are also particularly abundant in TMC-1 (see \citealt{Cernicharo2020b} and references therein).
The chemistry of this peculiar object produces a large abundance of
nearly saturated species, such as CH$_3$CHCH$_2$; this species may mostly be a typical molecule of hot cores
\citep{Marcelino2007}. The first
polar benzenic ring, C$_6$H$_5$CN \citep{McGuire2018}, was also detected in this object, while benzene itself has only been detected to date
towards post-asymptotic giant branch objects \citep{Cernicharo2001}. Hence, it has been a surprising result to see
that, with an almost dominant carbon chemistry, O-bearing carbon chains such as CCO \citep{Ohishi1991},
C$_3$O \citep{Matthews1984}, HC$_5$O, and HC$_7$O are also
produced in TMC-1 \citep{Cordiner2017,McGuire2017}. The formation path
for HC$_5$O and HC$_7$O is still a mystery, as is the reason for the non-detection of HC$_3$O,
HC$_4$O, and HC$_6$O in the same cloud. The O-bearing carbon chain HC$_2$O has, however, been observed in
cold dense clouds \citep{Agundez2015a}.

All long carbon chains containing oxygen that have been observed so far in interstellar clouds are 
neutral. Cationic species related to these oxygen-bearing neutral species
are thought to play an important role in the synthesis of different neutral molecules in cold dense
clouds. Moreover, it has been suggested that some of them may be sufficiently long-lived to be
abundant \citep{Petrie1993}, although to date no such cation has been observed. In general,
the abundance of polyatomic cations in cold interstellar clouds is relatively low because they react
quickly with electrons. In addition to the widespread HCO$^+$ and
N$_2$H$^+$, the other polyatomic cations detected in cold interstellar clouds are HCS$^+$ \citep{Thaddeus1981},
HCNH$^+$ \citep{Schilke1991},
HC$_3$NH$^+$ \citep{Kawaguchi1994},
HCO$_2^+$ (Turner et al. 1999; Sakai et al. 2008),
NH$_3$D$^+$ \citep{Cernicharo2013},
NCCNH$^+$ \citep{Agundez2015b}, H$_2$COH$^+$ \citep{Bacmann2016}, H$_2$NCO$^+$ \citep{Marcelino2018},
and HC$_5$NH$^+$ \citep{Marcelino2020}. These species are the protonated forms of abundant neutral species.
The abundance ratio between a protonated molecule and its neutral counterpart, [MH$^+$]/[M], is
sensitive to the degree of ionization and thus to various physical parameters of the cloud,
as well as the formation and destruction rates of the cation \citep{Agundez2015b}. The protonated form is mainly
formed by proton transfer to the neutral and destroyed by dissociative recombination with electrons. It is
interesting to note that both chemical models and observations suggest a trend in which the abundance ratio [MH$^+$]/[M]
increases with the increasing proton affinity of M \citep{Agundez2015b}.

In this letter, we report the detection of four harmonically related lines that belong to a
molecule with a $^1\Sigma$ ground electronic state towards the cold dark core TMC-1. From microwave laboratory experiments supported by 
\emph{ab initio} calculations, we conclude that the carrier is HC$_3$O$^+$, the protonated
form of C$_3$O. We present a detailed observational study of the most relevant O-bearing species in this
cloud and discuss these results in the context of state-of-the-art chemical models.

\section{Observations}

New receivers, built
as part of the Nanocosmos
project\footnote{\texttt{https://nanocosmos.iff.csic.es/}} and installed at the Yebes 40m radio telescope, were used
for the observations of TMC-1. The Q-band receiver consists of two HEMT (high electron mobility transistor)
cold amplifiers covering the
31.0-50.3 GHz band with horizontal and vertical polarizations. Receiver temperatures
vary from 22 K at 32 GHz to 42 K at 50 GHz. The spectrometers are $2\times8\times2.5$ GHz FFTs (Fast Fourier Transform)
with a spectral resolution of 38.1 kHz, which provides full coverage of the Q-band in both polarizations.
The main beam efficiency varies from 0.6 at 32 GHz to 0.43 at 50 GHz.

The observations leading to the line survey in Q-band towards TMC-1
($\alpha_{J2000}=4^{\rm h} 41^{\rm  m} 41.9^{\rm s}$ and $\delta_{J2000}=+25^\circ 41' 27.0''$)
were performed over several sessions carried out
between November 2019 and February 2020. The observing procedure was
frequency switching with a frequency throw of 10\,MHz. The nominal spectral
resolution of 38.1 kHz was used for the final spectra. The sensitivity varies along the
Q-band between 1 and 3 mK, which is a considerable improvement over previous line surveys in the 31-50 GHz frequency range
\citep{Kaifu2004}.

\begin{figure}[]
\centering
\includegraphics[width=0.85\columnwidth,angle=0]{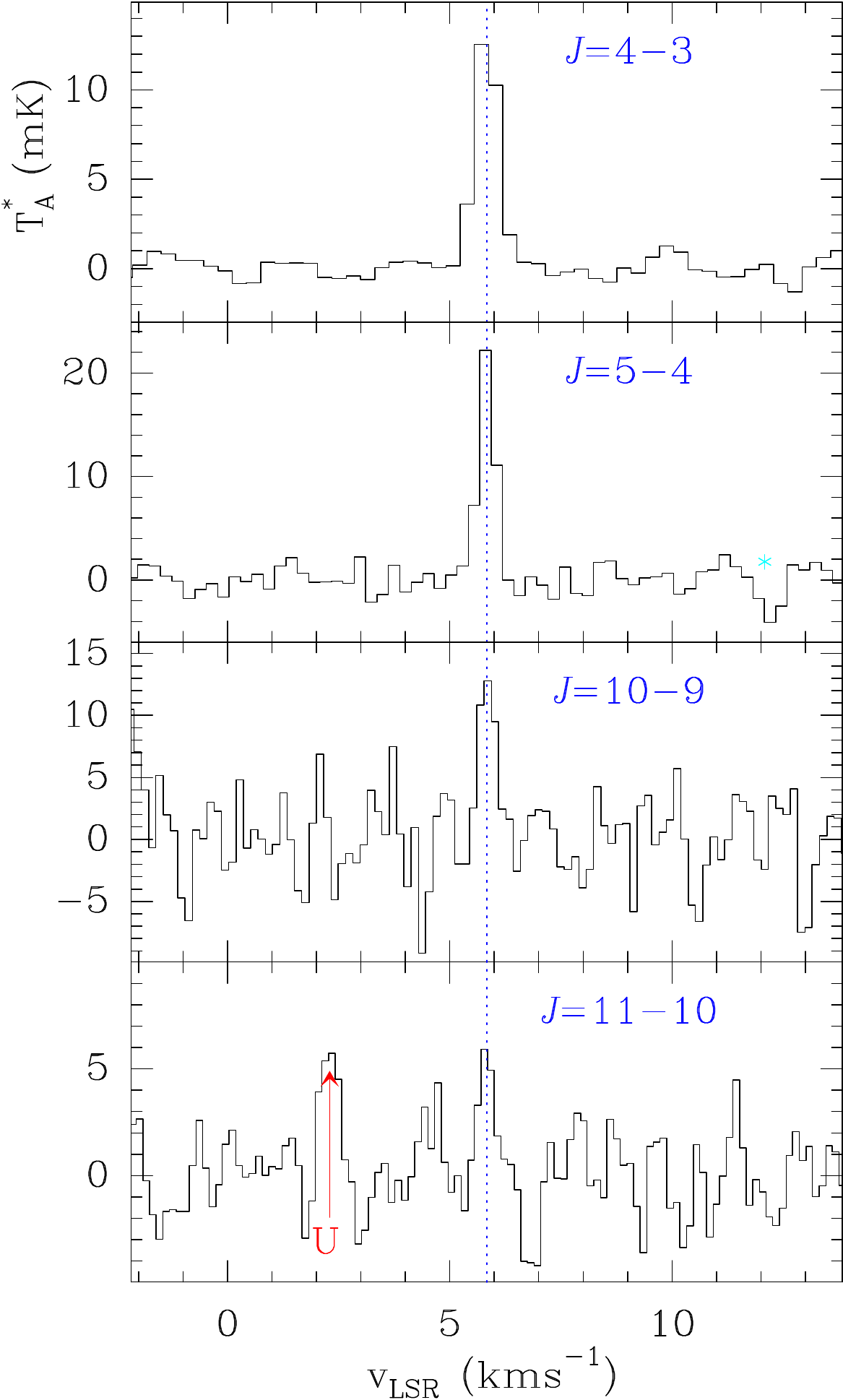}
\caption{Observed lines of the new molecule found in the 31-50 GHz domain towards TMC-1.
The abscissa corresponds to the local standard of rest velocity
in km s$^{-1}$. Frequencies and intensities for the observed lines are given in Table \ref{tab_hc3o+}.
The ordinate is the antenna temperature corrected for atmospheric and telescope losses in mK. The {\textit J=19-18}
line is only detected at a 3.5$\sigma$ level. Spectral resolution is 38.1 kHz.}
\label{fig_hc3o+}
\end{figure}
The IRAM 30m data come from a line survey performed towards TMC-1 and B1. These observations
have been described by \citet{Marcelino2007} and \cite{Cernicharo2013}.

The antenna temperature intensity scale
for the two telescopes used in 
this work was calibrated using two absorbers at different temperatures and the
atmospheric transmission model ATM \citep{Cernicharo1985, Pardo2001}.
Calibration uncertainties were adopted to be 10~\%.
All the data have been analysed using the GILDAS package\footnote{\texttt{http://www.iram.fr/IRAMFR/GILDAS}}.

\begin{table}
\tiny
\caption{Observed line parameters for the new molecule in TMC-1.}
\label{tab_hc3o+}
\centering
\begin{tabular}{{cccccc}}
\hline
{\textit J$_u$-J$_l$}& $\nu_{obs}^a$& $\nu_{o}-\nu_{c}^b$  & $\int$T$_A^*$dv $^c$ & $\Delta$v$^d$ & T$_A^*$\\
                     &  (MHz)       &     (kHz)            & (mK km\,s$^{-1}$)      & (km\,s$^{-1}$)  & (mK)\\
\hline
%2-1$^e$&17842.339&      &            &              &     \\
%3-2$^e$&26763.470&      &            &              &     \\
4-3    &35684.590&  -0.6& 9.1$\pm$0.5& 0.63$\pm$0.03& 13.8\\
5-4    &44605.748&  +4.4&10.7$\pm$0.6& 0.45$\pm$0.03& 22.5\\
10-9   &89209.744& -11.3& 6.4$\pm$1.0& 0.44$\pm$0.09& 13.8\\
11-10  &98130.286&   8.0& 2.4$\pm$0.7& 0.40$\pm$0.10&  5.8\\
\hline
\end{tabular}
\tablefoot{\\
        \tablefoottext{a}{Observed frequencies adopting a v$_{LSR}$ of 5.83 km s$^{-1}$ for TMC-1. The uncertainty is 10 kHz
    for all the lines.}\\
        \tablefoottext{b}{Observed minus calculated frequencies in kHz.}\\
        \tablefoottext{c}{Integrated line intensity in mK km\,s$^{-1}$. }\\
        \tablefoottext{d}{Line width at half intensity derived by fitting a Gaussian line profile to the observed
     transitions (in km\,s$^{-1}$).}
}
\end{table}
\normalsize
\section{Results}
\label{sec:results}
Most of the weak lines found in our survey of TMC-1 can be assigned to known species and their isotopologues,
and only a few remain unidentified (Marcelino et al., in preparation). As previously mentioned, the level of sensitivity has
been increased by a factor of 5-10 with respect to previous line surveys performed with other telescopes at these
frequencies \citep{Kaifu2004}. Frequencies for the unknown lines were derived by assuming a local standard of rest 
velocity of 5.83 km s$^{-1}$, a value that was derived from the observed transitions of HC$_5$N and its isotopologues in 
our line survey \citep{Cernicharo2020a}. Our new data towards TMC-1 have allowed us to detect the anions
C$_3$N$^-$ and C$_5$N$^-$ \citep{Cernicharo2020a}, the cation HC$_5$NH$^+$ \citep{Marcelino2020}, and the isomer HC$_4$NC 
of HC$_5$N \citep{Cernicharo2020b}, in addition to dozens of already known molecules.

The assignment of the observed features was done using the CDMS catalogue \citep{Muller2005}
and the MADEX code \citep{Cernicharo2012}. Within the still unidentified features in our survey, we found two lines
with a harmonic relation of 5:4 between them. The precision of this ratio is better than 2$\times$10$^{-5}$.
Taking into account the density of lines in TMC-1, the possibility that the harmonic relation between these two lines is
the result of a fortuitous
agreement is very small. Moreover, we explored our data at 3mm \citep{Marcelino2007} and found two additional
lines, the $J$=10-9 line at 89209.749 MHz and the $J$=11-10 transition at 98130.267 MHz.
The frequency relation 4:5:10:11 between the four lines strongly suggests that the carrier is a linear molecule
with a $^1\Sigma$ ground electronic state. The observed lines are shown in
Fig. \ref{fig_hc3o+}, and the derived
line parameters are given in Table \ref{tab_hc3o+}.

For a linear molecule in a $^1\Sigma$ electronic state, the frequencies of its rotational transitions follow the
standard expression $\nu(J\rightarrow$$J-1)$=2$B_0 J$ - 4$D_0 J^3$.
By fitting the frequencies of the lines  given in Table \ref{tab_hc3o+},
we derive $B_0$= 4460.58989$\pm$0.00096 MHz and
$D_0$= 51.06$\pm$0.47 Hz. The standard deviation of the fit is 10.3 kHz.

\subsection{Potential carriers of the series of lines}

An important piece of information to know when searching for the carrier of the
lines is that the $J$=4-3 line does not show any evidence of hyperfine splitting. A molecule with a terminal
CN group will produce observable hyperfine splitting for its $J$=4-3 transition. Hence, if the molecule
contains a CN group, it should be an isocyano species (-NC terminal group). The typical hyperfine quadrupole 
splitting of a terminal NH group will also be unresolved for the $J$=4-3 transition.

In TMC-1, only polyatomic molecules containing H, C, N, O, and S have been found so far. In order to evaluate the
possible structure of the carrier, we can consider other species that have a similar rotational constant. Molecules such as
HC$_3$N ($B_0\sim$4549 MHz), HC$_3$$^{15}$N ($B_0\sim$4417 MHz), C$_4$H and C$_4$D ($B_0\sim$4758 MHz and
4416.4 MHz, respectively),
%; but with a $^2\Sigma$ ground electronic state)
the radical HCCCO (quasi-linear
with (B+C)/2$\sim$4533 MHz),
% but a $^2\Pi$ ground electronic state), 
NCCNH$^+$ ($B_0\sim$4438 MHz), or H$_2$C$_4$
($(B+C)/2\sim$4466 MHz) all provide a strong indication of the presence of
four heavy atoms of C, N, and/or O, as well as a hydrogen or deuterium atom. Species containing sulphur, such as
HCCS, 
%($B_0\sim$5875 MHz),
%and a $^2\Pi$ ground electronic state)
HCCCS,  
%($B_0\sim$2688 MHz), 
or HNCS, have rotational constants that are very different from the observed one. 
%The molecule HNCS has also a too large rotational constant.

In fact, despite being an asymmetrical molecule, the species that has a close match to our rotational constant 
is H$_2$C$_4$; its case has to be considered in detail as its series of $J_{0,J}$ transitions are in 
good harmonic relation. Its lines are particularly
prominent in TMC-1, T$_A^*\sim$0.2 K. Although the measured $(B+C)/2$ value is $\sim$6 MHz above the observed one, 
its isotopologues H$_2$C$^{13}$CCC and H$_2$CC$^{13}$CC could have a slightly lower $(B+C)/2$. Its four 
$^{13}$C isotopologues have been observed in the laboratory \citep{McCarthy2002}. Those corresponding to the second 
and third carbon in the molecule have $(B+C)/2\sim$4454.8 and 4443.8 MHz, respectively. Hence, 
they do not match our $B$ rotational constant. Moreover, taking into account the derived $^{12}$C/$^{13}$C 
abundance ratio in TMC-1 of $\sim$90 \citep{Cernicharo2020b,Taniguchi2016}, the expected intensities
for the four $^{13}$C isotopologues of H$_2$C$_4$ are ten times weaker than the intensity of the observed lines 
(see Table \ref{tab_hc3o+} and Fig. \ref{fig_hc3o+}). Finally, the rotational constant of the HDCCCC 
isotopologue is smaller than that of our series of lines \citep{Kim2005}.

The cationic species derived from the protonation of CCCO, HCCCO$^+$, was calculated
by \citet{Botschwina1989} to have a $^1\Sigma$ ground electronic state with $B_0$=4454$\pm10$ MHz.
Recently, \citet{Thorwirth2020}
performed high-level \emph{ab initio} calculations and derived a rotational constant $B_0$$\sim$4460.4 MHz, 
which matches our observed value very well. This species may be formed from the reaction of C$_3$O with 
cations such as H$_3$$^+$ and HCO$^+$. The proton affinity of C$_3$O was calculated by \cite{Botschwina1989} 
to be quite high, 885$\pm$5 kJ mol$^{-1}$. From the observed trend of increasing protonated-to-neutral abundance 
ratios with increasing proton affinity \citep{Agundez2015b}, a high 
abundance ratio HC$_3$O$^+$/C$_3$O is expected.

Although HC$_3$O$^+$ seems
to be the best candidate, we have to explore other possible species derived from the isomers of NCCNH$^+$, which
also has a very close rotational constant of 4438.012 MHz \citep{Gottlieb2000}. In particular, CNCNH$^+$, the protonated
form of CNCN, which has also been found in space \citep{Agundez2018}, could also be a good candidate.
Other isomers with the proton on the terminal carbon atom are possible candidates as well (see Table \ref{abini_full}).
The protonated forms of the isomers of
HC$_3$N -- HCCNCH$^+$ and HNCCCH$^+$ -- could also have rotational constants around the observed value. The neutral
isomers of cyanoacetylene have been observed towards TMC-1 and are abundant (see \citealt{Cernicharo2020b} and
references therein), and the protonated form of HC$_3$N has also been detected in this source (see \citealt{Marcelino2020}
and references therein).

\begin{table}
\caption{Rotational constants and electric dipole moments of potential candidates.}
\label{abini_full}
\centering
\begin{tabular}{{|l|c|c|c|}}
\hline
\hline
Molecule  & $B_0$ & $D$ & $\mu$$^a$ \\
          & (MHz) &  (kHz) & (D)   \\
\hline
New species$^b$ & 4460.6     & 0.510 &       \\
\hline
HC$_3$O$^+$$^c$  & 4460.5    & 0.469 & 3.41  \\
HCNCN$^+$$^d$    & 4792.8    & 0.554 & 8.33  \\
HNCNC$^+$$^d$    & 4899.1    & 0.511 & 3.96  \\
HCNNC$^+$$^d$    & 5210.8    & 0.611 & 6.17  \\
HCCNCH$^+$$^d$   & 4646.4    & 0.458 & 3.45  \\
HNCCCH$^+$$^d$   & 4327.0    & 0.395 & 1.15  \\
HCCCO$^-$$^d$    & 4246.8    & 0.311 & 0.31  \\
\hline
\end{tabular}
\tablefoot{\\
    \tablefoottext{a}{Dipole moment calculated at the CCSD(T)-F12/cc-pCVTZ-F12 level of theory.}
    \tablefoottext{b}{Values derived from the frequencies observed in TMC-1 (see Sect. \ref{sec:results})}.
    \tablefoottext{c}{$B_0$ and $D$ values scaled by the ratio Exp/Calc. of the corresponding parameter HC$_3$N species.
    (See Table \ref{abini})}.
    \tablefoottext{d}{All the parameters for these species have been calculated at the CCSD(T)-F12/cc-pCVTZ-F12
     level of theory. No corrections using scaling factors have been applied.}
}
\end{table}

\begin{table*}
\caption{Theoretical values for the spectroscopic parameters of HC$_3$O$^+$ (all in MHz).}
\label{abini}
\centering
\begin{tabular}{{lccccc}}
\hline
\hline
&\multicolumn{2}{c}{HC$_3$N}&\multicolumn{3}{c}{HC$_3$O$^+$} \\
\cmidrule(lr){2-3} \cmidrule(lr){4-6}
Parameter & Calc.\tablefootmark{a} & Exp.\tablefootmark{b} & Calc.\tablefootmark{a} & Scaled.\tablefootmark{c} & Scaled.\tablefootmark{d} \\
\hline
$B_e$             &  4549.911 &                        &   4457.383        &               &                \\
Vib-Rot. Corr.    &   4.899   &                        &    1.630          &               &                \\
$B_0$             & 4544.215  &      4549.058558(40)   &   4455.753        &     4460.502  &    4460.443    \\
$D$ x 10$^{-3}$   & 0.501     &       0.5442223(91)    &    0.432          &     0.469     &       0.471    \\
\hline
\end{tabular}
\tablefoot{\\
        \tablefoottext{a}{This work; the $B_0$ rotational constant has been estimated using the $B_e$
value calculated at the CCSD(T)-F12/cc-pCVTZ-F12 level of theory and corrected with vibration-rotation
interaction estimated at the MP2/cc-pVTZ level of theory. The centrifugal distortion constant has been calculated
at the MP2/cc-pVTZ level of theory.} 
\tablefoottext{b}{\citet{Thorwirth2000}.} 
\tablefoottext{c}{This work; scaled by the ratio Exp/Calc. of the corresponding parameter for the HC$_3$N species}.
\tablefoottext{d}{\citet{Thorwirth2020}.
}
}
\end{table*}

\subsection{Quantum chemical calculations and assignment to HC$_3$O$^+$}
\label{ab_initio}
In order to obtain precise geometries and spectroscopic molecular parameters that help in the
assignment of the observed lines, we carried out high-level \emph{ab initio} calculations for
all the species mentioned above. Table \ref{abini_full} shows the results from the geometry optimization
calculations carried out at the CCSD(T)-F12/cc-pCVTZ-F12 level of theory
\citep{Raghavachari1989,Adler2007,Knizia2009,Hill2010a,Hill2010b}. This method has
been proven to be suitable for accurately reproducing the molecular geometry of analogue
molecules \citep{Cernicharo2019,Marcelino2020}. As can be seen, HCNCN$^+$, HNCNC$^+$,
HCNNC$^+$, and HCCNCH$^+$ all have rotational constants much larger than those derived from the
frequencies observed in TMC-1. Hence, they can be excluded as carriers of the observed lines.

The $B_e$  value obtained for HC$_3$O$^+$ is 4457.383 MHz (see Table \ref{abini}), which is very
close to that derived for the new molecule. To obtain a more precise value for the rotational
constant for HC$_3$O$^+$, we should estimate $B_0$ by the zero-point vibrational contribution to
the rotational constant and then calibrate this value using an experimental over theoretical ratio as
scaling factor 
for analogue molecular species. At this point, two options are possible: C$_3$O, which shares practically
the same molecular structure, or HC$_3$N, which is an isoelectronic species of HC$_3$O$^+$. Structural
calculations at the CCSD(T)-F12/cc-pCVTZ-F12 level of theory give a $B_e$ value for C$_3$O of 4795.4 MHz,
not very close to the experimental one of 4810.885 (33) \citep{Brown1983}. This discrepancy may be attributed
to the floppy nature of C$_3$O, for which large zero-point vibrational contributions are expected.
\citep{Botschwina1991} For this reason, C$_3$O is not a good reference system to calibrate the HC$_3$O$^+$
calculations, and, as such, we used HC$_3$N for this purpose. Table \ref{abini} shows the results for our calculations
for HC$_3$N and those for HC$_3$O$^+$. After adding zero-point vibrational contribution to the
$B_e$ rotational constant and scaling by the ratio Experimental/Calculated for HC$_3$N, we obtained a $B_0$ for HC$_3$O$^+$
of 4460.502 MHz, which agrees perfectly with that derived from the TMC-1 lines. The centrifugal distortion
value, obtained in the same manner but at the MP2/cc-pVTZ level of theory, is 0.469 kHz, which is compatible with
that obtained from the fit of the lines. The results of our calculations are in agreement with those
obtained by \citet{Thorwirth2020}, which are also shown in Table \ref{abini}.
Finally, another potential candidate is HC$_3$O$^-$. However, our calculations indicate a rotational constant
$\sim$210 MHz below the observed one and a very low permanent dipole moment (see Table \ref{abini}). Hence, from the arguments provided in the previous section and our calculations, we conclude that the
best candidate for the carrier of the observed lines is HC$_3$O$^+$.

\subsection{Laboratory detection of HC$_3$O$^+$}
\begin{figure}
\includegraphics[scale=0.52]{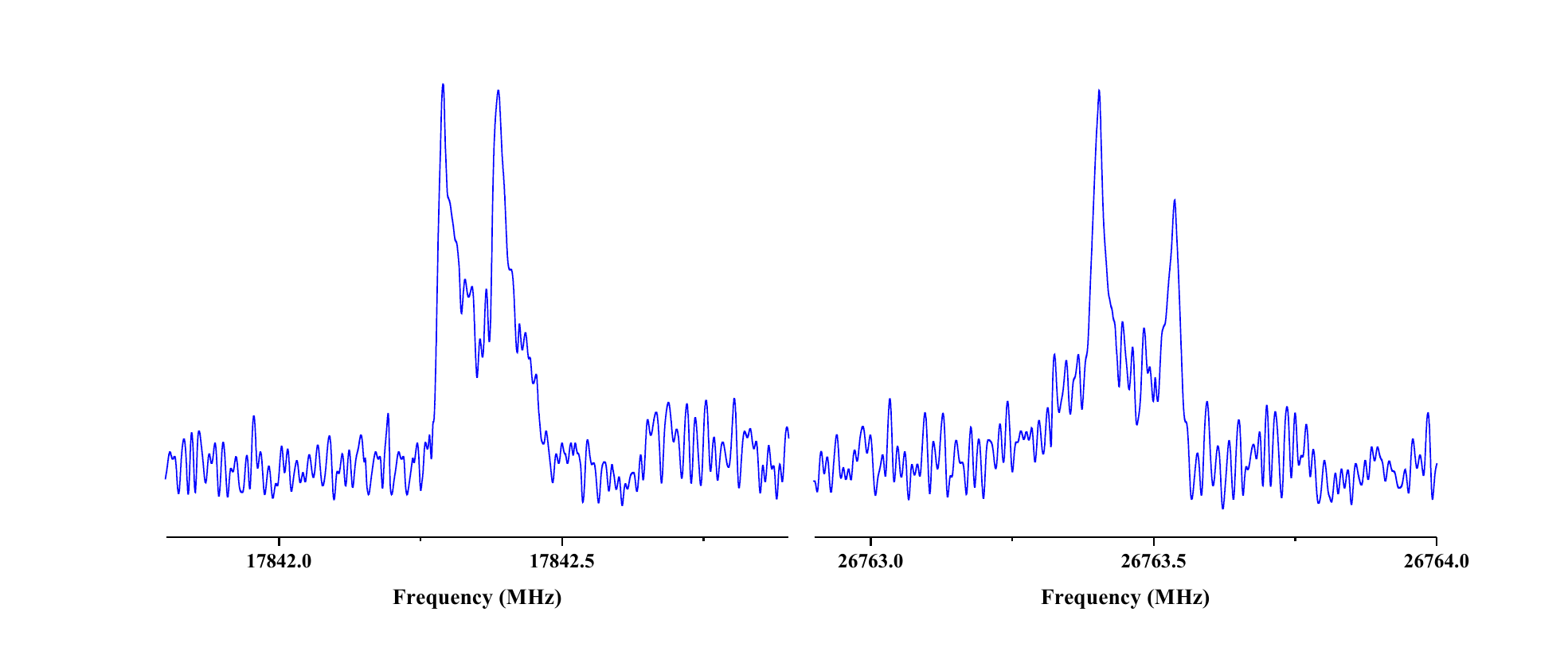}
\centering
\caption{\label{FTMW_spe} FTMW spectra of HC$_3$O$^+$ showing the 2-1  and 3-2 rotational transitions at 17842.3387 and 26763.4695 MHz, respectively. The spectra were
achieved via 20000 shots of accumulation at a repetition rate of 10 Hz. The coaxial arrangement of the adiabatic
expansion and the resonator axis produces an instrumental Doppler doubling. The resonance frequencies are
calculated as the average of the two Doppler components.}
\end{figure}

In order to confirm our assignment of the astrophysical lines to HC$_3$O$^+$, we
measured its microwave spectrum using a Balle-Flygare-type Fourier transform microwave (FTMW) spectrometer
combined with a pulsed discharge nozzle \citep{Endo1997,Cabezas2016}, which has been used in the past to characterize
other highly reactive molecules. The reactive transient species, HC$_3$O$^+$, was produced in a supersonic expansion by
a pulsed electric discharge of a gas mixture of C$_2$H$_2$ (0.15\%), CO (0.8\%), and H$_2$ (1.0\%) diluted in Ne, and  with the application of a voltage
of 800 V through the throat of the nozzle source. We searched for the $J$=2-1 and 3-2 rotational transitions of
HC$_3$O$^+$, predicting their frequencies from the rotational constants derived in TMC-1.
These frequency regions were scanned and two lines were observed, within a few kilohertz of the predicted frequencies, at
17842.3387 and 26763.4749 MHz with an uncertainty of 5 kHz (see Fig. \ref{FTMW_spe}).
The following experimental
results confirm that these lines belong to a transient species: (i) they disappear in the absence of electric discharge
and (ii) the lines disappear when one of the reactants (C$_2$H$_2$ or CO) is removed from the gas mixture. No
more lines at lower or higher frequencies ($J_u$-$J_l$ = 1-0 and 4-3) could be observed due to the spectrum weakness
and the worse performance of the spectrometer at those frequencies.

We definitively conclude that the carrier of the observed lines is protonated tricarbon monoxide (HC$_3$O$^+$). By merging the
laboratory and astrophysical data, we derive

$B_0$= 4460.58896$\pm$0.00057 MHz,

$D_0$= 50.64$\pm$0.30 Hz,

\noindent
which are the recommended constants for
predicting the rotational spectrum of HC$_3$O$^+$. The standard deviation of the merged fit is 9.1 kHz. Frequency
predictions have uncertainties $\le$10 kHz for transitions below 100 GHz and $\le$100 kHz for 
those in the range 100-180 GHz (see Table \ref{predicted}).

\begin{figure}[]
\centering
\includegraphics[width=\columnwidth,angle=0]{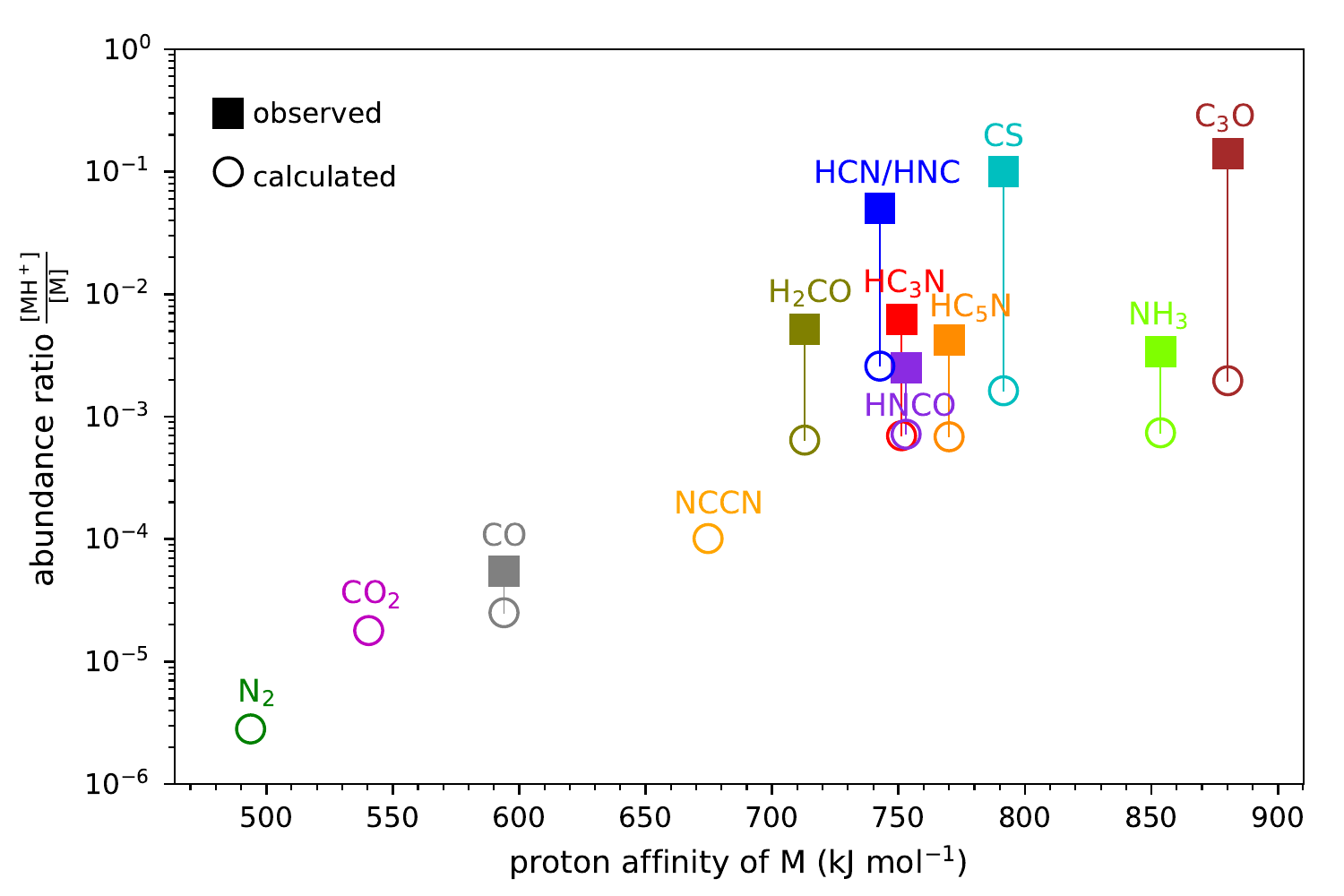}
\caption{Observed and calculated abundance ratios [MH$^+$]/[M] in cold dense clouds as a function of the proton affinity of
the neutral M. They are based on \cite{Agundez2015b}, with updates for H$_2$COH$^+$ from \cite{Bacmann2016}, for
H$_2$NCO$^+$ from \cite{Marcelino2018}, for HC$_5$NH$^+$ from \cite{Marcelino2020}, and for HC$_3$O$^+$ from this study.}
\label{fig_ratios}
\end{figure}

\section{Discussion}
The detection of HC$_3$O$^+$ in TMC-1 provides a further clue regarding the chemistry of protonated molecules
in cold dense clouds. \citet{Agundez2015b} find a trend in which the abundance ratio MH$^+$/M increases with the increasing proton affinity of the neutral M. 
The C$_3$O species has a very large proton affinity, $\sim$885$\pm$5 kJ mol$^{-1}$
\citep{Botschwina1989}. In fact, it is one of the largest proton affinities among abundant molecules in cold
dark clouds. The observed abundance ratio $N$(HC$_3$O$^+$)/$N$(C$_3$O)=0.14
(see Appendix \ref{O-bearing} and Table \ref{col_densities})
 fits well with the expected
value for a molecule with such a large proton affinity, according to the trend shown in Fig. \ref{fig_ratios}.
However, chemical model calculations similar to those presented in \cite{Agundez2015b} predict an abundance ratio
HC$_3$O$^+$/C$_3$O of only $2\times10^{-3}$ at steady state, which is almost two orders of magnitude below the
observed value. Part of the discrepancy may stem from the fact that rate constants for the main reactions of
the formation (proton transfer from H$_3$O$^+$, HCO$^+$, and H$_3^+$) and destruction (dissociative recombination with
electrons) of HC$_3$O$^+$ are not known, and thus chemical networks like UMIST {\small RATE12} \citep{McElroy2013} and
KIDA {\small \texttt{kida.uva.2014}} \citep{Wakelam2015a} adopt estimates for them. These estimates, however,
should not be drastically different from the real ones. In general, protonated-to-neutral abundance ratios are
underestimated by chemical models, probably because chemical networks miss important formation routes for
the cation (see \citealt{Agundez2015b} and Fig. \ref{fig_ratios}). In the case of HC$_3$O$^+$, additional
formation routes may be provided by reactions of the ions C$_3$H$_2$$^+$ and linear C$_3$H$_3$$^+$ with O atoms,
which is currently not considered in the aforementioned reaction networks. The reaction 
between C$_4$H$_2$$^+$ and O atoms (e.g. \citealt{Tenenbaum2006}) is only an efficient route to 
HC$_3$O$^+$  at very early times ($<10^5$ yr), while reactions of C$_3$H$^+$ with O$_2$, 
H$_2$O, and CO$_2$ \citep{Herbst1984} are not a major route to HC$_3$O$^+$ according to 
the chemical model. The sulphur analogue HC$_3$S$^+$ is a good candidate for detection 
given that C$_3$S has a very high proton affinity (933 kJ mol$^{-1}$) and is around five 
times more abundant than C$_3$O. We searched for it in our line survey using the \textit{ab 
initio} calculations of \citet{Thorwirth2020} and derive an upper limit to its 
column density of 3$\times$10$^{11}$ cm$^{-2}$, which implies an abundance 
ratio $N$(HC$_3$S$^+$)/$N$(C$_3$S) $<$ 0.04. The lower MH$^+$/M ratio of C$_3$S compared to C$_3$O 
probably stems from the fact that routes to HC$_3$O$^+$ involving O atoms can be efficient 
because of the high abundance of neutral O atoms, while the sulphur analogous routes would be 
less efficient because atomic sulphur is expected to be preferably in ionized rather than neutral form.

In Table~\ref{col_densities}, we present a comprehensive list of abundances derived from our data for O-bearing
molecules in TMC-1. The chemistry of some of these species has already been discussed in detail,
for example, in \cite{Agundez2015a} and \cite{Wakelam2015b} for HCCO, in \cite{Loison2016} for the isomers $c$-H$_2$C$_3$O
and HCCCHO, and in \cite{McGuire2017} and \cite{Cordiner2017} for HC$_5$O and HC$_7$O. These last
O-bearing carbon chains deserve some discussion. They represent a new class of interstellar molecules and their chemistry
is not yet well understood. \cite{Cordiner2012} proposed that HC$_n$O molecules may directly
form through reactions of hydrocarbon anions C$_n$H$^-$ with O atoms in a process of associative electron
detachment. This mechanism, however, overestimates the abundances of HC$_6$O, C$_6$O, and C$_7$O by one to two 
orders of magnitude \citep{Cordiner2017}. \cite{McGuire2017} proposed that HC$_n$O molecules could form through 
 radiative association reactions of C$_n$H$_2$$^+$ and C$_n$H$_3$$^+$ ions with CO followed by dissociative 
recombination with electrons. This mechanism explains the non-detection of HC$_6$O in terms of a low reactivity 
between C$_5$H$_2$$^+$ and CO \citep{Adams1989}, although it overestimates the abundance of HC$_4$O by almost
two orders of magnitude. The large abundance found for HC$_3$O$^+$ suggests that reactions between hydrocarbon 
ions and atomic oxygen probably participate in the growth of these long O-bearing carbon chains.

\begin{acknowledgements}

The Spanish authors thank Ministerio de Ciencia e Innovaci\'on for funding
support through projects AYA2016-75066-C2-1-P, PID2019-107115GB-C21. We also thank ERC for funding through grant
ERC-2013-Syg-610256-NANOCOSMOS. MA and CB thanks MICIN
for grants RyC-2014-16277 and FJCI-2016-27983, respectively. Y. Endo thanks Ministry of Science
and Technology of Taiwan through grant MOST108-2113-M-009-25.
\end{acknowledgements}

\normalsize

\begin{appendix}

\section{Observational data for O-bearing species in TMC1}
\label{O-bearing}
\begin{figure}[]
\centering
\includegraphics[width=0.88\columnwidth,angle=0]{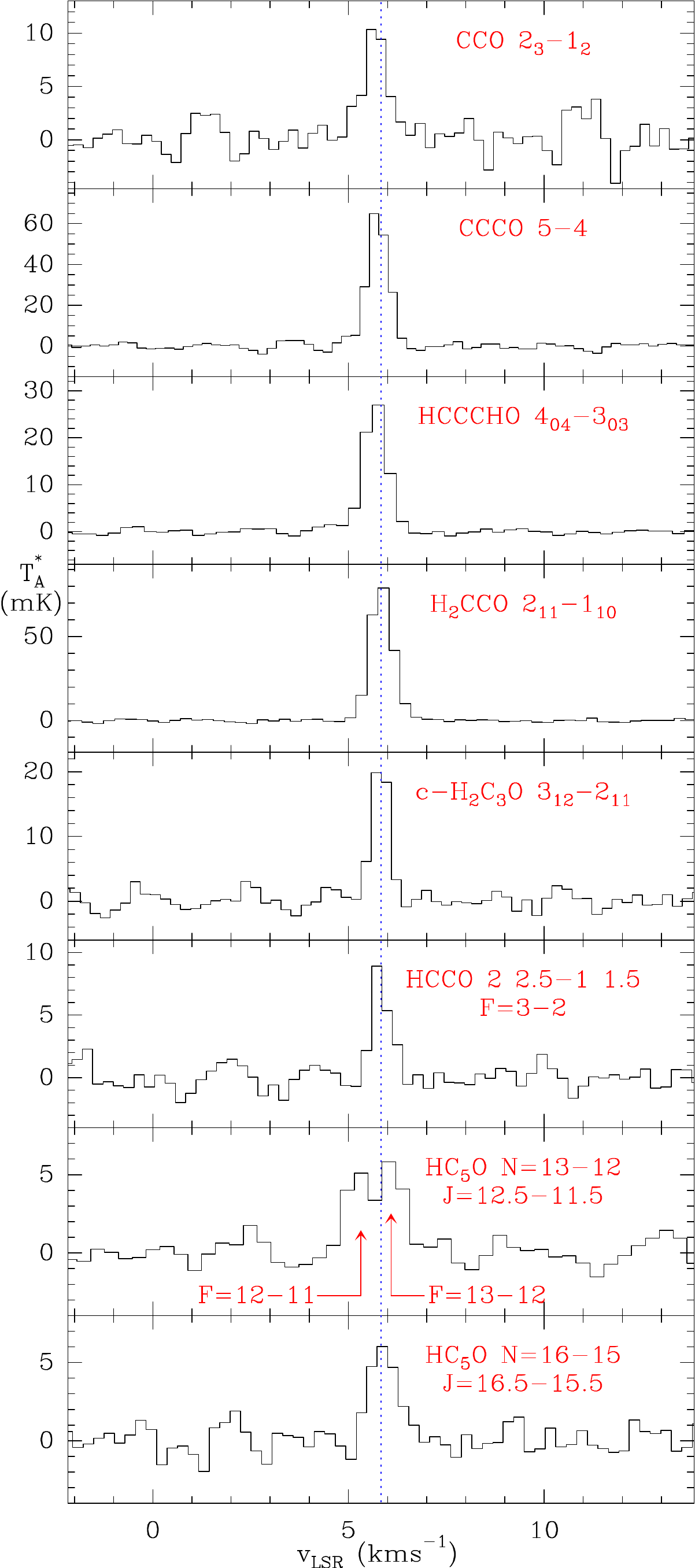}
\caption{Selected lines of O-bearing species observed towards TMC1 in the 31-50 GHz domain.
The abscissa corresponds to the local standard of rest velocity
in km s$^{-1}$. Frequencies and intensities for the observed lines are given in Table \ref{obs_O_parameters}.
The ordinate is the antenna temperature corrected for atmospheric and telescope losses in mK. Spectral
resolution is 38.1 kHz.}
\label{obs_O_lines}
\end{figure}

In order to study the chemistry of O-bearing species in TMC-1, we
explored the lines arising from these species  in our line survey. The derived
line parameters are given in Table \ref{obs_O_parameters}, and some selected
lines are shown in Fig. \ref{obs_O_lines}.
Line parameters were derived by fitting a Gaussian line profile to the
observed features. Rest frequencies were adopted from the MADEX code
\citep{Cernicharo2012}. In order to derive column densities, we assumed
that all molecular species are thermalized at a temperature of 10 K. Main
beam efficiency (see Sect. 2) and source dilution in the beam were
applied to all the observations. For the source dilution, we assumed
a source radius of 40$''$ \citep{Fosse2001}, that is to say, the source practically fills the
main beam of the telescope at all observed frequencies. The derived column
densities of detected O-bearing species in our survey were obtained using
the MADEX code and are given in Table \ref{col_densities}.
The 1$\sigma$ sensitivity of the survey is $\sim$0.6-0.7 mK and 0.7-2.0 mK below and
above 40 GHz, respectively. Hence, the data
allow us to derive very sensitive 3$\sigma$ upper limits for the column
density of many O-bearing species, which are given in Table \ref{col_densities}.
It is worth noting that although many lines of HC$_5$O
are detected in our survey, only upper limits have been obtained for HC$_7$O lines,
which has been reported previously by \cite{McGuire2017} and \cite{Cordiner2017} using line stacking.

Table \ref{predicted} provides the frequencies, uncertainties, upper energy
levels, Einstein coefficients, and degeneracies for transitions up to $J$=30-29.
This information is provided with the intent of facilitating the search for HC$_3$O$^+$ in other sources and with other
instruments. 

\begin{table}
\caption{Column densities for relevant O-bearing species in TMC-1.}
\label{col_densities}
\centering
\tiny
\begin{tabular}{{|l|c|c|c|l|}}
\hline
Molecule        & $N^a$                      & $X^b$                    & $N$(CH$_3$OH)/$N$ & \\
\hline
HCCCO$^+$       &  2.1(2)$\times$10$^{11}$   & 2.1$\times$10$^{-11}$ & 229      &  \\
HOCO$^+$        &  4.0(2)$\times$10$^{11}$   & 4.0$\times$10$^{-11}$ & 120      &  \\
H$_2$COH$^+$    & $\le$3.0$\times$10$^{11}$  &$\le$3.0$\times$10$^{-11}$& $\ge$160 &  \\
H$_2$NCO$^+$    & $\le$4.0$\times$10$^{10}$  &$\le$4.0$\times$10$^{-12}$& $\ge$1200&  \\
HNCO            & 1.3(1)$\times$10$^{13}$    & 1.3$\times$10$^{-9}$  & 3.7      &     \\
HCNO            & 7.0(3)$\times$10$^{10}$    & 7.0$\times$10$^{-12}$ & 686      &     \\
HOCN            & 1.1(2)$\times$10$^{11}$    & 1.1$\times$10$^{-11}$ & 437      &     \\
CH$_3$OH        &  4.8(3)$\times$10$^{13}$   & 4.8$\times$10$^{-9}$  & 1        & $A+E$\\
CCO             &  1.5(3)$\times$10$^{12}$   & 1.5$\times$10$^{-10}$ & 32       &  \\
CCCO            &  1.5(2)$\times$10$^{12}$   & 1.5$\times$10$^{-10}$ & 32       &  \\
CCCCO           & $\le$1.2$\times$10$^{11}$  &$\le$1.2$\times$10$^{-11}$& $\ge$400 &  \\
CCCCCO          & $\le$4.0$\times$10$^{10}$  &$\le$4.0$\times$10$^{-12}$& $\ge$1200&  \\
HCOOH           &  1.4(2)$\times$10$^{12}$   & 1.4$\times$10$^{-10}$ & 34       &  \\
HCCO            &  1.0(2)$\times$10$^{12}$   & 1.0$\times$10$^{-10}$ & 48       &  \\
HCCCO           & $\le$2.0$\times$10$^{11}$  &$\le$2.0$\times$10$^{-11}$& $\ge$240 &  \\
HCCCCO          & $\le$3.0$\times$10$^{11}$  &$\le$3.0$\times$10$^{-11}$& $\ge$160 &  \\
HCCCCCO         &  1.8(2)$\times$10$^{12}$   & 1.8$\times$10$^{-10}$ & 27       &  \\
HCCCCCCO        & $\le$7.0$\times$10$^{11}$  &$\le$7.0$\times$10$^{-11}$& $\ge$59  &  \\
HCCCCCCCO       & $\le$9.0$\times$10$^{11}$  &$\le$9.0$\times$10$^{-11}$& $\ge$53  &  \\
H$_2$CCO        &  1.4(2)$\times$10$^{13}$   & 1.4$\times$10$^{-9}$  & 3.4      &  $o+p$ \\ % p 2.9(2)E12
H$_2$CCCO       & $\le$1.1$\times$10$^{11}$  &$\le$1.1$\times$10$^{-11}$& $\ge$436 &  \\
c-H$_2$C$_3$O   &  4.0(2)$\times$10$^{11}$   & 4.0$\times$10$^{-11}$ & 120      & $o+p$ \\ %o 3.0E11 p 1.0E11
HCCCHO          &  2.0(2)$\times$10$^{12}$   & 2.0$\times$10$^{-10}$ & 240      &  \\
H$_2$CCCCO      & $\le$1.2$\times$10$^{11}$  &$\le$1.2$\times$10$^{-11}$& $\ge$400 & $o$ \\
CH$_3$CHO       &  3.5(2)$\times$10$^{12}$   & 3.5$\times$10$^{-10}$ & 14       &  $A+E$ \\ %A 1.8E12 E 1.7E12
NH$_2$CHO       & $\le$5.0$\times$10$^{10}$  &$\le$5.0$\times$10$^{-12}$& $\ge$960 &  \\
HCCCCOH         & $\le$1.0$\times$10$^{11}$  &$\le$1.0$\times$10$^{-11}$& $\ge$480 &  \\
\hline
\end{tabular}
\tablefoot{\\
        \tablefoottext{a}{Column density in cm$^{-2}$. Upper limits correspond to 3$\sigma$ values.}\\
        \tablefoottext{b}{Relative abundance to H$_2$ assuming a total column density of
    molecular hydrogen of 10$^{22}$ cm$^{-2}$ \citep{Cernicharo1987}.}\\
}
\end{table}
\normalsize

\onecolumn
\begin{longtable}{|l|c|c|c|c|c|c|}
%\centering
\caption[]{Derived line parameters for O-bearing species in TMC-1.
\label{obs_O_parameters}}\\
\hline
Molecule        & Transition        & $\nu_{rest}$$^a$   & $\int$T$_A^*$dv $^b$ & v$_{LSR}$$^c$ & $\Delta$v$^d$ & T$_A^*$$^e$\\
                &                   &    (MHz)           &  (mK km\,s$^{-1}$)   & (km\,s$^{-1}$)& (km\,s$^{-1}$)& (mK)\\
\hline
\endfirsthead
\caption{continued.}\\
\hline
 Molecule        & Transition        & $\nu_{rest}$$^a$   & $\int$T$_A^*$dv $^b$ & v$_{LSR}$$^c$ & $\Delta$v$^d$ & T$_A^*$\\
                &                   &    (MHz)           &  (mK km\,s$^{-1}$)   & (km\,s$^{-1}$)& (km\,s$^{-1}$)& (mK)\\
\hline
\endhead
\hline
\endfoot
\hline
\endlastfoot
\hline
%CCO N=1.5(3)E12
CCO             & 1$_1$-0$_1$       & 32623.449$\pm$0.007& 5.1$\pm$0.7 & 5.57(05)& 1.09(13)& 4.5\\ %OK
CCO             & 2$_1$-1$_1$       & 32738.613$\pm$0.005& 1.6$\pm$0.7 & 6.33(18)& 1.46(34)& 1.0\\ %demasiado debil predicted identical to previous
CCO             & 2$_3$-1$_2$       & 45826.734$\pm$0.002& 8.1$\pm$0.9 & 5.69(04)& 0.73(11)& 10.0\\ %demasiado fuerte
CCO             & 2$_2$-1$_1$       & 46182.187$\pm$0.002& 2.5$\pm$1.0 & 5.73(17)& 0.89(31)& 2.7\\ % muy debil predicted factor 2 weaker than previous
\hline
%N(C3O) N=1.5(2)E12 cm-2
CCCO            &     4-3           & 38486.891$\pm$0.001&43.5$\pm$1.0 & 5.74(01)& 0.63(07)&64.5\\
CCCO            &     5-4           & 48108.474$\pm$0.001&44.2$\pm$1.0 & 5.75(01)& 0.63(14)&66.4\\
\hline
%HNCO N=1.3(1)E13
HNCO          &2$_{02}$-1$_{01}$ 1-1& 43962.007$\pm$0.030 & 20.1$\pm$1.0 &5.76(01)& 0.62(02)&35.1\\
HNCO          &2$_{02}$-1$_{01}$ 1-2& 43962.641$\pm$0.006 &              &        &         &$\le$2.7\\
HNCO          &2$_{02}$-1$_{01}$ 3-2& 43963.000$\pm$0.030 &178.8$\pm$1.0 &5.76(01)& 0.68(01)& 247.8\\
HNCO          &2$_{02}$-1$_{01}$ 2-1& 43963.000$\pm$0.030 &              &        &         &      \\
HNCO          &2$_{02}$-1$_{01}$ 1-0& 43963.626$\pm$0.030 & 26.1$\pm$2.0 &6.35(02)& 0.60(04)& 41.2\\
HNCO          &2$_{02}$-1$_{01}$ 2-2& 43963.626$\pm$0.030 & 22.9$\pm$2.0 &5.66(03)& 0.64(05)& 33.6\\
\hline
%HCNO N=7E10
HCNO          & 2-1                 & 45876.069$\pm$0.002 & 7.1$\pm$1.0  &5.79(08)& 0.98(16)&  6.9\\
\hline
%HOCN N=1.1(2)E11
HOCN          & 2$_{02}$-1$_{01}$   & 41950.836$\pm$0.001 & 13.6$\pm$1.0  &5.55(02)& 0.82(03)& 15.6\\
\hline
%HCOOH N=1.4(3)E12 N_cis<1.2E11
HCOOH           & 2$_{12}$-1$_{11}$ & 43303.705$\pm$0.001& 3.4$\pm$0.5 & 6.10(09)& 0.61(10)& 5.2\\
HCOOH           & 2$_{02}$-1$_{01}$ & 44911.734$\pm$0.001& 6.7$\pm$0.5 & 5.82(05)& 0.79(10)& 7.9\\
HCOOH           & 2$_{11}$-1$_{10}$ & 46581.220$\pm$0.001& 2.1$\pm$0.5 & 5.95(14)& 0.70(22)& 2.8\\
\hline
%HCCCHO N=2.0(0.3)E12
HCCCHO          & 4$_{14}$-3$_{13}$ & 36648.266$\pm$0.006& 7.4$\pm$0.5 & 5.76(02)& 0.78(05)&  9.0\\
HCCCHO          & 4$_{04}$-3$_{03}$ & 37290.136$\pm$0.006&21.0$\pm$0.5 & 5.64(01)& 0.70(02)& 28.1\\
HCCCHO          & 4$_{13}$-3$_{12}$ & 37954.572$\pm$0.006&10.4$\pm$0.5 & 5.59(01)& 0.74(04)& 13.2\\
HCCCHO          & 5$_{15}$-4$_{14}$ & 45807.708$\pm$0.007& 3.5$\pm$0.7 & 5.73(05)& 0.46(10)&  7.3\\
HCCCHO          & 5$_{05}$-4$_{04}$ & 46602.868$\pm$0.007&25.5$\pm$0.5 & 5.64(01)& 0.65(02)&  3.6\\
HCCCHO          & 5$_{14}$-4$_{13}$ & 47440.427$\pm$0.007& 6.0$\pm$0.7 & 5.61(04)& 0.55(09)& 10.3\\
\hline
%A-CH3CHO N=1.8E12 E-CH3CHO n=1.8E12
A-CH$_3$CHO    & 4$_{ 04}$-3$_{ 13}$& 32709.214$\pm$0.002&             &         &        & $\le$1.8 \\
A-CH$_3$CHO    & 2$_{ 12}$-1$_{ 11}$& 37464.204$\pm$0.001&12.1$\pm$1.0 & 5.84(02)&0.91(06)& 12.5     \\
A-CH$_3$CHO    & 2$_{ 02}$-1$_{ 01}$& 38512.079$\pm$0.001&31.3$\pm$0.6 & 5.85(01)&0.82(01)& 35.7     \\
A-CH$_3$CHO    & 2$_{ 11}$-1$_{ 10}$& 39594.289$\pm$0.001&14.0$\pm$1.0 & 5.83(02)&0.79(05)& 16.6     \\
A-CH$_3$CHO    & 1$_{ 10}$-1$_{ 01}$& 47820.620$\pm$0.002& 4.3$\pm$0.6 & 5.76(08)&0.65(13)&  6.1     \\
A-CH$_3$CHO    & 2$_{ 11}$-2$_{ 02}$& 48902.831$\pm$0.002& 4.0$\pm$0.6 & 5.76(07)&0.66(13)&  5.6     \\
E-CH$_3$CHO    & 4$_{+04}$-3$_{-13}$& 33236.468$\pm$0.002&             &         &        & $\le$2.0 \\
E-CH$_3$CHO    & 2$_{-12}$-1$_{+10}$& 35837.312$\pm$0.003& 1.3$\pm$0.6 & 5.89(09)&0.62(20)&  2.0     \\
E-CH$_3$CHO    & 2$_{-12}$-1$_{-11}$& 37686.932$\pm$0.002& 9.7$\pm$1.0 & 5.88(02)&0.75(05)& 12.1     \\
E-CH$_3$CHO    & 2$_{+02}$-1$_{+01}$& 38506.035$\pm$0.001&33.7$\pm$1.0 & 5.85(01)&0.82(01)& 38.5     \\
E-CH$_3$CHO    & 2$_{+11}$-1$_{+10}$& 39362.537$\pm$0.002&13.8$\pm$1.0 & 5.82(02)&0.84(04)& 15.4     \\
E-CH$_3$CHO    & 2$_{+11}$-1$_{-11}$& 41212.157$\pm$0.003& 2.5$\pm$0.8 & 6.07(09)&0.83(16)&  2.8     \\
E-CH$_3$CHO    & 1$_{-11}$-1$_{+01}$& 45897.347$\pm$0.003&             &         &        & $\le$3.0 \\
E-CH$_3$CHO    & 1$_{+10}$-1$_{+01}$& 47746.968$\pm$0.002& 2.5$\pm$0.8 & 6.17(22)&0.82(23)&  2.8     \\
E-CH$_3$CHO    & 2$_{+11}$-2$_{+02}$& 48603.469$\pm$0.002& 6.8$\pm$1.0 & 5.88(05)&0.75(11)&  8.6     \\
\hline
%o-H2CCO N(o-H2CCO)=1.1(1)E13 N(p-H2CCO)=2.9(1)E12
o-H$_2$CCO      & 2$_{12}$-1$_{11}$ & 40039.017$\pm$0.001&57.8$\pm$0.7 & 5.77(01)&0.68(01)& 79.5     \\
o-H$_2$CCO      & 2$_{11}$-1$_{10}$ & 40793.842$\pm$0.001&59.4$\pm$0.7 & 5.79(01)&0.68(01)& 81.9     \\
p-H$_2$CCO      & 2$_{02}$-1$_{01}$ & 40417.950$\pm$0.001&37.9$\pm$0.7 & 5.78(01)&0.64(02)& 55.9     \\
\hline
%o-c-H2C3O N=3.0(2)E11 %p-c-H2C3O N=1.0(4)E11
o-c-H$_2$C$_3$O & 3$_{13}$-2$_{12}$ & 39956.733$\pm$0.004& 14.9$\pm$0.8& 5.81(01)&0.65(03)& 21.5     \\
o-c-H$_2$C$_3$O & 3$_{12}$-2$_{11}$ & 44587.397$\pm$0.004& 12.4$\pm$0.8& 5.82(01)&0.51(03)& 22.8     \\
p-c-H$_2$C$_3$O & 3$_{03}$-2$_{02}$ & 42031.939$\pm$0.004&  6.4$\pm$0.8& 5.78(03)&0.68(07)&  8.9     \\
p-c-H$_2$C$_3$O & 3$_{22}$-2$_{21}$ & 42316.187$\pm$0.003&  1.7$\pm$0.7& 5.78(02)&0.27(10)&  6.0     \\
p-c-H$_2$C$_3$O & 3$_{21}$-2$_{20}$ & 42601.246$\pm$0.003&  2.4$\pm$0.7& 5.94(03)&0.48(09)&  4.6     \\
\hline
%N(HCCO)=1E12
HCCO            & 2,2.5-1,1.5 F=3-2 & 43317.673$\pm$0.006&  4.9$\pm$0.8& 5.80(03)&0.51(08)&  8.9     \\
HCCO            & 2,2.5-1,1.5 F=2-1 & 43321.150$\pm$0.006&  3.8$\pm$0.8& 5.70(05)&0.67(15)&  5.4     \\
HCCO            & 2,1.5-1,0.5 F=2-1 & 43329.543$\pm$0.006&  3.8$\pm$0.8& 5.67(06)&0.64(15)&  5.5     \\
HCCO            & 2,1.5-1,0.5 F=1-0 & 43335.462$\pm$0.004&  4.2$\pm$0.8& 5.69(08)&0.97(17)&  4.1     \\
\hline
%N(HC5O)=1.8E12 TROT=10  mu=2.16
HC$_5$O&13  1 12.5-12 -1 11.5 13-12    &32267.964$\pm$0.002& 3.7$\pm$0.9  & 5.52(06)&0.63(09)&  5.6      \\
HC$_5$O&13  1 12.5-12 -1 11.5 12-11    &32268.049$\pm$0.002& 4.8$\pm$1.0  & 5.65(07)&0.73(15)&  6.0      \\
HC$_5$O&13 -1 12.5-12  1 11.5 13-12    &32271.760$\pm$0.002& 2.7$\pm$1.0  & 5.59(08)&0.62(14)&  3.9      \\
HC$_5$O&13 -1 12.5-12  1 11.5 12-11    &32271.848$\pm$0.002& 4.3$\pm$1.0  & 5.62(06)&0.76(15)&  5.4      \\
HC$_5$O&14 -1 13.5-13  1 12.5 14-13    &34849.461$\pm$0.002& 1.9$\pm$0.7  & 5.58(02)&0.33(16)&  5.0      \\
HC$_5$O&14 -1 13.5-13  1 12.5 13-12    &34849.540$\pm$0.002& 4.5$\pm$1.0  & 5.67(03)&0.81(08)&  5.3      \\
HC$_5$O&14  1 13.5-13 -1 12.5 14-13    &34853.387$\pm$0.002& 4.5$\pm$1.0  & 5.69(17)&0.74(14)&  5.7      \\
HC$_5$O&14  1 13.5-13 -1 12.5 13-12    &34853.469$\pm$0.002& 3.1$\pm$0.9  & 5.63(14)&0.54(16)&  5.4      \\
HC$_5$O&15  1 14.5-14 -1 13.5 15-14$^f$&37430.982$\pm$0.003& 6.3$\pm$1.0  & 5.79(03)&0.93(05)&  6.4      \\
HC$_5$O&15 -1 14.5-14  1 13.5 15-14$^f$&37435.050$\pm$0.003& 6.2$\pm$1.0  & 5.79(04)&0.99(09)&  5.9      \\
HC$_5$O&15 -1 15.5-15  1 14.5 16-15$^f$&40012.451$\pm$0.004& 5.3$\pm$1.0  & 5.80(06)&1.00(12)&  4.9      \\
HC$_5$O&15  1 15.5-15 -1 14.5 16-15$^f$&40016.668$\pm$0.004& 7.1$\pm$1.0  & 5.86(04)&1.02(10)&  6.5      \\
HC$_5$O&16  1 16.5-15 -1 15.5 17-16$^f$&42593.906$\pm$0.004& 5.2$\pm$1.0  & 5.70(05)&0.77(09)&  6.3      \\
HC$_5$O&16 -1 16.5-15  1 15.5 17-16$^f$&42598.284$\pm$0.004& 5.4$\pm$1.0  & 5.89(04)&0.82(09)&  6.2      \\
HC$_5$O&17 -1 17.5-16  1 16.5 18-17$^f$&45175.346$\pm$0.005& 5.4$\pm$1.0  & 5.75(11)&1.05(19)&  4.8      \\
HC$_5$O&17  1 17.5-16 -1 16.5 18-17$^f$&45179.895$\pm$0.005& 6.3$\pm$1.0  & 5.80(08)&1.06(17)&  5.6      \\
HC$_5$O&18  1 18.5-17 -1 17.5 19-18$^f$&47756.772$\pm$0.007&              &         &        &  $\le$5.1 \\
HC$_5$O&18 -1 18.5-17  1 17.5 19-18$^f$&47761.501$\pm$0.007&              &         &        &  $\le$5.1 \\
\hline
\end{longtable}
\tablefoot{\\
        \tablefoottext{a}{Rest frequencies and uncertainties from MADEX \citep{Cernicharo2012}.}\\
        \tablefoottext{b}{Integrated line intensity in mK kms$^{-1}$.}\\
        \tablefoottext{c}{Local standard of rest velocity of the observed emission kms$^{-1}$.}\\
        \tablefoottext{d}{Line width at half intensity derived by fitting a Gaussian line profile to the observed
     transitions (in kms$^{-1}$).}\\
        \tablefoottext{e}{Upper limits to the antenna temperature correspond to 3$\sigma$ values.}\\
        \tablefoottext{f}{Average of two hyperfine components.}\\
}

\newpage
\begin{table}
\begin{center}
\caption{Predicted frequencies of HC$_3$O$^+$.}
\label{predicted}
\centering
\tiny
\begin{tabular}{{|c|c|c|c|c|c|}}
\hline
Transition& $\nu$              & E$_{up}$ & A$_{ij}$   &S$_{ij}$& g$_u$\\
          &  (MHz)             &  (K)     & (s$^{-1}$)  &        &      \\
\hline
          1$\rightarrow$ 0    &  8921.175$\pm$0.001&   0.4&2.928 10$^{-08}$&  1&  3\\
 2$\rightarrow$ 1    & 17842.338$\pm$0.003&   1.3&2.810 10$^{-07}$&  2&  5\\
 3$\rightarrow$ 2    & 26763.477$\pm$0.004&   2.6&1.016 10$^{-06}$&  3&  7\\
 4$\rightarrow$ 3    & 35684.579$\pm$0.005&   4.3&2.498 10$^{-06}$&  4&  9\\
 5$\rightarrow$ 4    & 44605.633$\pm$0.006&   6.4&4.990 10$^{-06}$&  5& 11\\
 6$\rightarrow$ 5    & 53526.627$\pm$0.006&   9.0&8.756 10$^{-06}$&  6& 13\\
 7$\rightarrow$ 6    & 62447.547$\pm$0.006&  12.0&1.406 10$^{-05}$&  7& 15\\
 8$\rightarrow$ 7    & 71368.383$\pm$0.006&  15.4&2.116 10$^{-05}$&  8& 17\\
 9$\rightarrow$ 8    & 80289.124$\pm$0.006&  19.3&3.033 10$^{-05}$&  9& 19\\
10$\rightarrow$ 9    & 89209.754$\pm$0.008&  23.5&4.182 10$^{-05}$& 10& 21\\
11$\rightarrow$10    & 98130.263$\pm$0.010&  28.3&5.590 10$^{-05}$& 11& 23\\
12$\rightarrow$11    &107050.638$\pm$0.014&  33.4&7.284 10$^{-05}$& 12& 25\\
13$\rightarrow$12    &115970.868$\pm$0.020&  39.0&9.290 10$^{-05}$& 13& 27\\
14$\rightarrow$13    &124890.942$\pm$0.027&  45.0&1.163 10$^{-04}$& 14& 29\\
15$\rightarrow$14    &133810.844$\pm$0.035&  51.4&1.434 10$^{-04}$& 15& 31\\
16$\rightarrow$15    &142730.566$\pm$0.045&  58.2&1.744 10$^{-04}$& 16& 33\\
17$\rightarrow$16    &151650.094$\pm$0.056&  65.5&2.095 10$^{-04}$& 17& 35\\
18$\rightarrow$17    &160569.416$\pm$0.069&  73.2&2.491 10$^{-04}$& 18& 37\\
19$\rightarrow$18    &169488.521$\pm$0.083&  81.3&2.934 10$^{-04}$& 19& 39\\
20$\rightarrow$19    &178407.395$\pm$0.099&  89.9&3.427 10$^{-04}$& 20& 41\\
21$\rightarrow$20    &187326.026$\pm$0.118&  98.9&3.971 10$^{-04}$& 21& 43\\
22$\rightarrow$21    &196244.404$\pm$0.138& 108.3&4.571 10$^{-04}$& 22& 45\\
23$\rightarrow$22    &205162.515$\pm$0.160& 118.2&5.227 10$^{-04}$& 23& 47\\
24$\rightarrow$23    &214080.348$\pm$0.185& 128.4&5.944 10$^{-04}$& 24& 49\\
25$\rightarrow$24    &222997.889$\pm$0.211& 139.1&6.724 10$^{-04}$& 25& 51\\
26$\rightarrow$25    &231915.129$\pm$0.240& 150.3&7.569 10$^{-04}$& 26& 53\\
27$\rightarrow$26    &240832.053$\pm$0.272& 161.8&8.482 10$^{-04}$& 27& 55\\
28$\rightarrow$27    &249748.650$\pm$0.306& 173.8&9.466 10$^{-04}$& 28& 57\\
29$\rightarrow$28    &258664.908$\pm$0.342& 186.2&1.052 10$^{-03}$& 29& 59\\
30$\rightarrow$29    &267580.815$\pm$0.381& 199.1&1.166 10$^{-03}$& 30& 61\\
\hline
\end{tabular}
\end{center}
\end{table}

\end{appendix}


\begin{thebibliography}{}
\tiny
\bibitem[Adler et al.(2007)]{Adler2007} Adler, T. B., Knizia, G., Werner. H.-J., 2007 J. Chem. Phys. 127, 221106
\bibitem[Adams et al.(1989)]{Adams1989} Adams, N. G., Smith, D., Giles, K., \& Herbst, E. 1989, \aap, 220, 269
\bibitem[Ag\'undez et al.(2015a)]{Agundez2015a} Ag\'undez, M., Cernicharo, J., \& Gu\'elin, M. 2015a, \aap, 577, L5
\bibitem[Ag\'undez et al.(2015b)]{Agundez2015b} Ag\'undez, M., Cernicharo, J., de Vicente, P., et al., 2015b, \aap, 579, L10
\bibitem[Ag\'undez et al.(2018)]{Agundez2018} Ag\'undez, M., Marcelino, N., \& Cernicharo, J. 2018, \apj, 861, L22
\bibitem[Bacmann et al.(2016)]{Bacmann2016} Bacmann, A., Garc\'ia-Garc\'ia, E., \& Faure, A. 2016, \aap, 588, L8
\bibitem[Botschwina(1989)]{Botschwina1989} Botschwina, P., 1989, J. Chem. Phys., 90, 4301
\bibitem[Botschwina \& Reisenauer (1991)]{Botschwina1991} Botschwina, P., Reisenauer, H. P., 1991, Chem. Phys. Lett., 183, 217.
\bibitem[Brown et al. (1983)]{Brown1983} Brown, R. D., Eastwood, F. W., Elmes, P. S., Godfrey, P. D., 1983 J. Am. Chem. Soc., 105, 6496.
\bibitem[Cabezas et al. (2016)]{Cabezas2016}Cabezas, C., Guillemin, J.-C., Endo, Y. 2016, J. Chem. Phys., 145, 184304.
\bibitem[Cernicharo(1985)]{Cernicharo1985} Cernicharo, J. 1985, Internal IRAM report (Granada: IRAM)
\bibitem[Cernicharo \& Gu\'elin(1987)]{Cernicharo1987} Cernicharo, J., \& Gu\'elin, M. 1987, \aap, 176, 299
\bibitem[Cernicharo et al.(2001)]{Cernicharo2001} Cernicharo, J., Heras, A.M, Tielens, A.G.G.M., et al., 2001, \apj, 546, L123
\bibitem[Cernicharo(2012)]{Cernicharo2012} Cernicharo, J., 2012, in ECLA 2011: Proc. of the European Conference on Laboratory Astrophysics,
EAS Publications Series, 2012, Ed.: C. Stehl, C. Joblin, \& L. d'Hendecourt (Cambridge: Cambridge Univ. Press),
251; \texttt{https://nanocosmos.iff.csic.es/?page$\_$id=1619}
\bibitem[Cernicharo et al.(2013)]{Cernicharo2013} Cernicharo, J., Tercero, B., Fuente, A., et al. 2013, \apj, 771, L10
%\bibitem[Cernicharo et al.(2018)]{Cernicharo2018} Cernicharo, J., Lefloch, B., Ag\'undez, M., et al., 2020, \apj, 853, L22
\bibitem[Cernicharo et al.(2019)]{Cernicharo2019} Cernicharo, J., Cabezas, C., Pardo, J.~R., et al., 2019, \aap, 630, L2
\bibitem[Cernicharo et al.(2020a)]{Cernicharo2020a} Cernicharo, J., et al., 2020a, \aap, 641, L9
\bibitem[Cernicharo et al.(2020b)]{Cernicharo2020b} Cernicharo, J., et al., 2020b, \aap, DOI:10.1051/0004-6361/202039274
\bibitem[Cordiner \& Charnely(2012)]{Cordiner2012} Cordiner, M. A. \& Chamley, S. B. 2012, \apj, 749, 120
\bibitem[Cordiner et al.(2017)]{Cordiner2017} Cordiner, M. A., Chamley, S. B., Kisiel, Z., et al., 2017, \apj, 850, 187
%\bibitem[Eichelberger et al.(2007)]{Eichelberger2007} Eichelberger, B., Snow, T. P., Barckholtz, C., \& Bierbaum, V. 2007, \apj, 667, 1283
\bibitem[Endo et al. (1997)]{Endo1997} Endo, Y., Kohguchi, H., Ohshima, Y. 1994, Faraday Discuss., 97, 341.
\bibitem[Foss\'e et al.(2001)]{Fosse2001} Foss\'e, D., Cernicharo, J., Gerin, M., Cox, P., 2001, \apj, 552, 168
\bibitem[Gottlieb et al.(2000)]{Gottlieb2000} Gottlieb, C.A., Aponi, A.J., McCarthy, M.C., Thaddeus, P., 2000, J. Chem. Phys., 113, 1910
\bibitem[Herbst et al. (1984)]{Herbst1984} Herbst, E., Smith, D., Adams, N.G., 1984, \aap, 138, L13
\bibitem[Hill et al.(2010a)]{Hill2010a} Hill, J. G., Mazumder, S., Peterson, K. A., 2010a, J. Chem. Phys., 132, 054108
\bibitem[Hill et al.(2010b)]{Hill2010b} Hill, J. G., Peterson, K. A., 2010b, Phys. Chem. Chem. Phys., 12, 10460
\bibitem[Kaifu et al.(2004)]{Kaifu2004} Kaifu, N., Ohishi, M., Kawaguchi, K., et al., 2004, PASJ, 56, 69
\bibitem[Kawaguchi et al.(1994)]{Kawaguchi1994}Kawaguchi, K., Kasai, Y., Ishikawa, S. et al., 1994, \apj, 420, L95
\bibitem[Kim \& Yamamoto(2005)]{Kim2005} Kim, E., Yamamoto, S., 2005, J. Mol. Spectrosc., 233, 93
\bibitem[Knizia et al.(2009)]{Knizia2009} Knizia, G., Adler, T. B., Werner. H.-J., 2009 J. Chem. Phys. 130, 054104
\bibitem[Loison et al.(2016)]{Loison2016} Loison, J.-C., Ag\'undez, M., Marcelino, N., et al. 2016, \mnras, 456, 4101
\bibitem[Matthews et al.(1984)]{Matthews1984} Matthews, H.E., Irvine, E., Friberg, F.M., et al., Nature, 310, 125
\bibitem[Marcelino et al.(2007)]{Marcelino2007} Marcelino, N., Cernicharo, J., Ag\'undez, M., et al. 2007, \apj, 665, L127
\bibitem[Marcelino et al.(2018)]{Marcelino2018} Marcelino, N., Ag\'undez, M., Cernicharo, J., et al. 2018, \aap, 612, L10
\bibitem[Marcelino et al.(2020)]{Marcelino2020} Marcelino, N., et al., 2020, \aap, DOI:10.1051/0004-6361/202039251
\bibitem[McCarthy \& Thaddeus(2002)]{McCarthy2002} McCarthy, M.C., Thaddeus, P., 2002, J. Mol. Spectrosc., 211, 235
\bibitem[McElroy et al.(2013)]{McElroy2013} McElroy, D., Walsh, C., Markwick, A. J., et al. 2013, \aap, 550, A36
\bibitem[McGuire et al.(2017)]{McGuire2017} McGuire, B.A., Burkhardt, M., Shingledecker, C.N., et al., 2017, \apj, 843, L28
\bibitem[McGuire et al.(2018)]{McGuire2018} McGuire, B.A., Burkhardt, M., Kalenskii, S., et al., 2018, Science, 359, 202
\bibitem[M\"uller et al.(2005)]{Muller2005} M\"uller, H.S.P., Schl\"oder, F., Stutzki, J., Winnewisser, G., 2005, \jmst, 742, 215
\bibitem[Ohishi et al.(1991)]{Ohishi1991} Ohishi, M., Suzuki, H., Ishikawa, S.-I., et al., 1991, \apj, 380, L39
\bibitem[Pardo et al.(2001)]{Pardo2001} Pardo, J.~R., Cernicharo, J., Serabyn, E. 2001, IEEE Trans. Antennas and Propagation, 49, 12
\bibitem[Petrie et al.(1993)]{Petrie1993} Petrie, S., Bettens, R. P. A., Freeman, C. G., \& McEwan, M. J. 1993, \mnras, 264, 862
\bibitem[Raghavachari et al.(1989)]{Raghavachari1989} Raghavachari, K., Trucks, G.~W., Pople, J.~A., Head-Gordon, M., 1989, Chem. Phys. Lett., 157, 479
\bibitem[Sakai et al.(2008)]{Sakai2008} Sakai, N., Sakai, T., Aikawa, Y., Yamamoto, S. 2008, \apj, 675, L89
\bibitem[Schilke et al.(1991)]{Schilke1991} Schilke, P., Walmsley, C. M., Millar, T. J., Henkel, C. 1991, \aap, 247, 487
\bibitem[Taniguchi et al. (2016)]{Taniguchi2016}Taniguchi, K., Ozeki, H., Saito, M., et al., 2016, \apj, 817, 147
\bibitem[Tenenbaum et al. (2006)]{Tenenbaum2006}Tenenbaum, E.D., Apponi, A.J., Ziurys, L.M., et al., 2006, \apj, 649, L17
\bibitem[Thaddeus et al.(1981)]{Thaddeus1981} Thaddeus, P., Gu\'elin, M., Linke, R. A. 1981, \apj, 246, L41
\bibitem[Thorwirth et al. (2000)]{Thorwirth2000}Thorwirth, S., M\"uller, H.~S.~P. Winnewisser, G., 2000, J. Mol. Spectrosc., 204, 133
\bibitem[Thorwirth et al. (2020)]{Thorwirth2020} Thorwirth, S., Harding, M.E., Asvany, O., et al., 2020, Mol. Phys., in press
\bibitem[Turner et al.(1999)]{Turner1999} Turner, B. E., Terzieva, R., Herbst, E. 1999, \apj, 518, 699
\bibitem[Wakelam et al.(2015a)]{Wakelam2015a} Wakelam, V., Loison, J.-C., Herbst, E., et al. 2015a, \apjs, 217, 20
\bibitem[Wakelam et al.(2015b)]{Wakelam2015b} Wakelam, V., Loison, J.-C., Hickson, K. M., \& Ruaud, M. 2015b, \mnras, 453, L48
\end{thebibliography}
\end{document}